\documentstyle[12pt]{article}
\begin{document}
\newcommand{\pl}{Phys.\ Lett.\ }
\newcommand{\npb}[1]{Nucl.\ Phys.\ {\bf B#1}\ }
\newcommand{\prd}[1]{Phys.\ Rev.\ {\bf D#1}\ }
\newcommand{\prl}{Phys.\ Rev.\ Lett.\ }
\newcommand{\hgg}{$ \phi \rightarrow \gamma \gamma $}
\newcommand{\hgz}{$ \phi \rightarrow \gamma Z^{0} $}
\newcommand{\sutwo}{$ SU(2)_{L} $}
\newcommand{\znot}{$ Z^{0} $}
\newcommand{\hff}{$\factor{a}{\Lambda} F_{\mu \nu} F^{\mu \nu} $}
\newcommand{\hfz}{$\factor{b}{\Lambda} F_{\mu \nu} Z^{\mu \nu} $}
\newcommand{\ogg}{$O_{\gamma \gamma}$}
\newcommand{\ogz}{$O_{\gamma Z}$}
\addtolength{\textheight}{0.4 in}

\title{Searching for the Scalar of the Strongly-Coupled Standard Model}

\author{Witold Skiba$^1$ \\
Center for Theoretical Physics \\
Massachusetts Institute of Technology \\
Cambridge, MA 02139}
\addtocounter{footnote}{1}
\footnotetext{This work is supported in
part by the U.S. Department of Energy under
cooperative agreement DE-FC02-94ER40818.\hfill}
\date{}

\maketitle

\vspace{-3.5in}
\rightline{ \begin{tabular}{l}
MIT-CTP-2417  \\
February 1995  \\
\end{tabular} }
\vspace{ 4.2in}

\begin{abstract}
We investigate decays of the scalar bound state present in the Abbott-Farhi
model. We show that decays with photons in the final state may have large
branching ratios. We also show that operators coupling the scalar particle
to two photons or to a photon and a \znot\ are not seriously constrained by
electroweak data, unlike other sectors of the Abbott-Farhi model.
\end{abstract}

\thispagestyle{empty}
\newpage
\setcounter{page}{1}

\section{Introduction}

Precise measurements of electroweak parameters \cite{LEP} have ruled out
or severely constrained many models. Although measurements with such a high
level of precision give some insight into high-energy phenomena, there
still exists a low-energy alternative to the Standard Model (SM) of
electroweak interactions, namely the Abbott-Farhi model \cite{AbbottFarhi}.
The Abbott-Farhi model, or the Strongly-Coupled Standard
Model (SCSM), has the same Lagrangian as the Minimal SM, however the \sutwo\
coupling constant becomes large at a characteristic scale higher than the
electroweak symmetry breaking scale in the SM. \sutwo\ interactions are
confining below that scale. Parameters of the scalar sector are adjusted so
that the \sutwo\ symmetry is not broken. Below the confinement scale the
particles are composite objects. The spectrum of the lightest particles
consists of fermions, the photon, three massive spin-one bosons and one
spinless particle.

Fermions are bound states of a fundamental fermion and a scalar.
Spin-one particles are bound states of two fundamental scalars; their
interactions resemble the interactions of the massive gauge bosons
of the SM. The composite scalar is a bound state of two fundamental scalars,
and its experimental signatures may be similar to those of the Higgs particle
of the Minimal SM. The composite particles do not carry any net \sutwo\
charge, just as the physical, hadronic states of QCD carry no net color.

At low energies, the interactions of the composite particles can be described
in terms of an effective Lagrangian. Neglecting higher-dimensional
operators and heavier, yet unobserved, particles, the low-energy couplings
of vector bosons to fermions are identical to those of the SM.
Thus, the low-energy effective theory of such particles, under certain
assumptions, reproduces physics of the SM \cite{AbbottFarhi,Claudson}.
Deviations from SM predictions test the magnitude of higher-dimensional
operators and the presence of exotic particles predicted by the SCSM:
excited fermions, excited vector bosons, as well as diquarks, dileptons
and leptoquarks. For a more detailed description of the model see
Ref.~\cite{Claudson}. However, there are some arguments based on lattice
studies and continuum field theory \cite{ALS,Hsu}  that the confining
phase of the \sutwo\ theory breaks chiral symmetries, and consequently there
cannot be light fermions present in the particle spectrum. It is not
entirely clear if those arguments apply to the Abbott-Farhi model, since
Yukawa couplings, strong $SU(3)$ and hypercharge $U(1)$ gauge
interactions were not taken into account.

Experiments before LEP did not seriously constrain the exotic sector of the
SCSM \cite{Wudka}.  A recent analysis of the model \cite{us} shows that
the SCSM is now so constrained that it may look unnatural. However, the SCSM
cannot be ruled out, as the exotic sector of the model does not influence
low-energy measurements (up to the \znot\ mass) in the limit of heavy
exotic particles and weak couplings. Another way to confirm or rule out
the SCSM could be the observation of new particles. It is possible
that, for some reason, the excited vector bosons analyzed in \cite{us}
are very heavy or else weakly coupled. Then the first new particle to be
observed could be the the lightest scalar particle. Can we distinguish
the SCSM scalar from the SM Higgs boson? If the SCSM is the true theory
of the weak interactions, at the phenomenological level, fermions and
vector particles are very similar to their SM counterparts.
It might well be that the total width of the SCSM scalar and its
partial widths of decays to fermions and massive vector bosons are
experimentally indistinguishable from those predicted for the SM Higgs
boson. However, the composite nature of the scalar particle may show up
in some decay channels of such a particle, especially decays with photons
in final state.

In this letter we analyze decays of the scalar particle into two
photons or into a photon and a \znot. In the SM, couplings of
the Higgs particle to two photons or one photon and the \znot\
are absent at the tree level, since the Higgs particle is a fundamental
neutral field. Such couplings first appear at the one-loop level. In the SCSM,
the photon can interact directly with the charged constituents of the scalar
bound state, making couplings considerably larger than the one-loop level
couplings in the SM. Large branching ratios of the decays \hgg\ and \hgz\
can be a clear signal for physics beyond the Minimal SM, and in particular,
a signal for the SCSM. We argue that operators responsible for decays of the
scalar into final states including photons may have very large magnitudes
in the SCSM. We estimate the numbers of such events that would be observable
at the LHC. Finally, we show that those operators are indeed not
significantly bounded by present electroweak data.

\section{The decays \hgg\ and \hgz}

The physical scalar particle of the SCSM is a bound state of fundamental
charged scalar fields, $\phi = \varphi^{*}_{\alpha} \varphi^{\alpha}$,
where $\varphi^{\alpha}$ is a fundamental scalar field, and $\alpha$ is
an \sutwo\ index. Fermion masses and fermion-scalar couplings for the
composite field both arise from Yukawa couplings of the elementary Higgs
in the underlying Lagrangian. Thus, the Yukawa couplings in the effective
theory are proportional to the fermion masses. The remaining couplings of
the scalar particle cannot be deduced without solving the confining dynamics
of the model. We therefore investigate two dimension-five operators that
could be large in the SCSM:
\begin{equation}
\label{operators}
O_{\gamma \gamma} = \frac{a}{\Lambda} \phi F_{\mu \nu} F^{\mu \nu}
 \  \  {\rm and} \ \
O_{\gamma Z} = \frac{b}{\Lambda} \phi F_{\mu \nu} Z^{\mu \nu},
\end{equation}
where $F_{\mu \nu} = \partial_{\mu} A_{\nu} - \partial_{\nu} A_{\mu}$,
$Z_{\mu \nu} = \partial_{\mu} Z_{\nu} - \partial_{\nu} Z_{\mu}$ (no
commutator), and $\Lambda$ is the characteristic scale of \sutwo\ interaction.
These operators are the lowest-dimensional operators consistent with
electromagnetic gauge invariance that give rise to the decays \hgg\ and
\hgz. Composite particles of the SCSM are \sutwo\ singlets, thus any operator
written in terms of composite fields is already \sutwo\ invariant.
By contrast, in the SM, an operator like $h F_{\mu \nu} Z^{\mu \nu}$ included
in the Lagrangian would spoil the gauge invariance.

Let us consider the operator \ogg\ and estimate its coefficient using
conventional dimensional analysis \cite{Georgi}. Operators are constructed
from dimensionless combinations like
$ e F_{\mu \nu} / \Lambda^2 $ and $ \phi / v $, and multiplied
by $v^2 \Lambda^2 $, where $e$ is the unit charge, $v$ is the electroweak
symmetry breaking scale $\approx 250$ GeV, and $\Lambda$ is of the order
$4 \pi v$. This estimate gives $O_{\gamma\gamma} = [ e^2 / (4 \pi \Lambda)]
\phi F_{\mu \nu} F^{\mu \nu}$. $\Lambda \approx 4 \pi v$ is the scale of
physics beyond the SM. In the SCSM, $\Lambda$ can be identified with the scale
of \sutwo\ interactions. Hence, the coefficient $a$ in Eq. (\ref{operators})
is equal to $\frac{e^2}{4 \pi}$. However, this estimate may be completely
irrelevant in case of the SCSM. Dimensional analysis is based on the
assumption that the derivative expansion of an effective Lagrangian works
as naively expected, i.e.\ higher-dimensional terms are small, comparable
to loop corrections of operators with fewer derivatives. The SCSM requires
an additional assumption \cite{Claudson} in order to reproduce the observed
weak interactions: the vector bosons $W$ and $Z$ have to be unnaturally light,
much lighter than the scale of \sutwo\ interactions. This assumption should
be contrasted with knowledge from QCD, where all particles (except Goldstone
bosons) are heavier than the scale $\Lambda_{\rm{QCD}}$.

It may indicate that the dynamics of the SCSM are quite complicated, and
consequently the derivative expansion of the Lagrangian may be unconventional.
This allows the possibility that in the SCSM the operators \ogg\ and \ogz\
could have much bigger magnitude than naively expected. The operator \ogg\
can arise from the diagram illustrated in Fig.~1. Photon fields couple
directly to charged scalar preons. The diagram shown in Fig.~1 is not a loop
diagram in the sense of perturbation theory because the couplings of the
$SU(2)_L$ bosons are non perturbative. The magnitude of the operator \ogg\
is proportional to the magnitude of preon wave function at the origin. We
assume that in the SCSM the magnitudes of the operators \ogg\ and \ogz\ are
not suppressed by non-perturbative effects, so the magnitudes can be as large
as $a=e^2$ and $b=e$. We will show in the next chapter that such large
coefficients are allowed by current experimental data.

Operators \ogg\ and \ogz\ induce decays of the scalar particle:
\hgg\ and \hgz. The resulting partial widths are then:
\begin{eqnarray}
\label{widthgg}
\Gamma(\phi \rightarrow \gamma \gamma) & = &
\frac{|a|^2}{\Lambda^2} \frac{m_{\phi}^3}{2 \pi}, \\
\label{widthgz}
\Gamma(\phi \rightarrow \gamma Z) & = &
\frac{|b|^2}{\Lambda^2} \frac{m_{\phi}^3}{8 \pi}
(1 - \frac{m_Z^2}{m_{\phi}^2})^3,
\end{eqnarray}
where $m_{\phi}$ is the mass of the scalar particle. As we mentioned before,
the magnitude of coupling between the scalar and a pair of fermions is
proportional to the fermion mass. Also the couplings of the scalar to the $W$
and $Z$ bosons can have magnitudes close to those of the SM Higgs particle,
since the couplings are described by dimension-three operators, which are not
suppressed by a large mass scale. However, the magnitudes of couplings
between the scalar and two photons or a photon and a \znot\ can be much
larger than in the SM. Thus the \hgg\ and \hgz\ decay modes are particularly
interesting to search for. By looking for these decays the SCSM scalar
can be discovered and, perhaps, distinguished from the SM Higgs boson.

In Fig.~2 we present the widths of the decays \hgg\ and \hgz\
computed from Eqs.~(\ref{widthgg}) and (\ref{widthgz}), with
$a=e^2$ and $b=e$. The widths are normalized to the total width
of the Higgs boson in the Minimal SM \cite{hunter}. We cannot
rigorously compute the total width of the scalar in the SCSM. The
normalization serves only as a reference point.
Clearly, branching fractions are much larger in the SCSM than in the SM,
where they are of the order $10^{-4}$ \cite{hunter}.

In Table 1 we present an estimate of the number of events for both decay
channels expected at the LHC in one year of running with an integrated
luminosity of $10^5 \ {\rm pb}^{-1}$ \cite{LHC}. We have again assumed that
the production cross section for the SCSM scalar is the same as for the
Minimal SM Higgs boson.

The widths of the dominant decay channels of the SCSM scalar are likely to be
similar to those of the SM Higgs particle. If the scalar is lighter than
about 150 GeV, decays into $b\bar{b}$ and $\tau^+\tau^-$ pairs will dominate.
In this mass range the scalar particle has a very small width, and can have
${\rm BR}(\phi \rightarrow \gamma \gamma)$ as big as 10\%. The \hgg\
channel is a clean and easy discovery mode of the SCSM scalar particle.
The number of events is so large that even if the mass of the scalar
coincides with the \znot\ mass, detection will be possible. Thus
the \hgg\ decay mode is very interesting to look for both at the LEP-II
and the LHC. The \hgz\ channel is interesting only in a very narrow mass
range from approximately 120 to 150 GeV, where the branching ratio can reach
a few percent.

If the scalar mass is above 150 GeV, the dominant decay mode will be
to $WW$ and $ZZ$ pairs (one of the bosons may be off the mass shell). The
partial width for the decays into a pair of vector bosons increases as
the cube of the scalar mass. With the heavy top quark \cite{top},
decays into $t\bar{t}$ pairs will have quite a small branching ratio,
and probably cannot be observed at the LHC. The most suitable channel
for the detection of the SCSM scalar is likely to be the mode
$\phi \rightarrow ZZ \rightarrow 4l$. However, the \hgg\ channel
could give a signal with larger statistical significance, depending on the
ability of the detector to suppress $\gamma \gamma$ background. For a
$\gamma \gamma$ invariant mass of more then 200 GeV there is very little
`irreducible' background from $q\bar{q},gg\rightarrow \gamma \gamma$
production. The main sources of background are misidentified $\pi^0$'s
and jets. Background rejection becomes a crucial feature of the detector
for detection of a scalar heavier than 300 GeV. The width of the scalar is
large, so background rejection may be more important than energy resolution.
The expected number of events from background at the proposed CMS experiment
corresponds to cross section $10 \ {\rm fb}$ per 1-GeV bin of $\gamma \gamma$
invariant mass \cite{CMS}. It will then be possible to observe the  \hgg\
signal with a statistical significance greater than 5 for a scalar lighter
than approximately 450 GeV.

The above discussion is based on the optimistic assumption that the
magnitudes of the operators \ogg\ and \ogz\ are as large as they possibly
can be. Should the actual magnitudes of these operators differ only
slightly from their magnitudes in the SM, then information about the decays
of the  lightest scalar into two photons or a photon and a $Z^0$ are not
enough to tell the SCSM from the SM or myriads of its extensions.

\section{Oblique corrections}

We are now going to show, as claimed earlier, that the operators \ogg\ and
\ogz\ are not sufficiently restricted by current precise electroweak
measurements to exclude their having large magnitudes. We use the formalism
of oblique corrections. Operators \ogg\ and \ogz\ contribute to self-energies
of the vector bosons, and they do not give any significant contribution to
four-fermion operators. Thus low-energy effects of these operators can be
conveniently expressed in terms of the parameters S, T and U \cite{Peskin}.
Self-energy diagrams are of the form shown in Fig.~3.
A one-loop calculation performed using dimensional regularization gives:
\begin{eqnarray}
\label{self}
\Pi_{\gamma \gamma}^{\mu \nu} & = &
\frac{|a|^2}{\pi^2 \Lambda^2} (q^2 g^{\mu \nu} - q^\mu q^\nu)
\left[\frac{m_{\phi}^2}{2} \left(D(m_\phi^2) + \frac{3}{2}\right)
+ O(q^2)\right], \nonumber \\
\Pi_{\gamma \gamma}^{\mu \nu} & = &
\frac{|b|^2}{4 \pi^2 \Lambda^2} (q^2 g^{\mu \nu} - q^\mu q^\nu)
\left[\frac{m_{\phi}^2 + m_Z^2}{2} \left(D(m_\phi^2) + \frac{3}{2}\right)
+ O(q^2)\right], \\
\Pi_{Z Z}^{\mu \nu} & = &
\frac{|b|^2}{4 \pi^2 \Lambda^2} (q^2 g^{\mu \nu} - q^\mu q^\nu)
\left[\frac{m_{\phi}^2}{2} \left(D(m_\phi^2) + \frac{3}{2}\right)
+ O(q^2)\right], \nonumber
\end{eqnarray}
where $D(m^2)=\frac{1}{2 \epsilon} - \gamma - \log(\frac{m^2}{4 \pi \mu^2})$.
We assume that the operators \ogg\ and \ogz\ are independent, i.e.\ there
is no accidental cancellation between their effects. The S, T and U
parameters are UV convergent if the self-energy counterterms are related
in the following way \cite{Golden}:
\begin{eqnarray}
\delta\Pi_{WW} = \delta\Pi_{W}, & \ \ &
\delta\Pi_{ZZ} = c_W^2 \delta\Pi_{W} + s_W^2 \delta\Pi_{B}, \nonumber \\
\delta\Pi_{\gamma Z} = c_W s_W(\delta\Pi_{W} - \delta\Pi_{B}), & \ \  &
\delta\Pi_{ZZ} = s_W \delta\Pi_{W} + c_W^2 \delta\Pi_{B}, \nonumber
\end{eqnarray}
where $\delta$ indicates a divergent counterterm, and $s_W$ and $c_W$ are,
respectively, sine and cosine of the Weinberg angle. In our case these
relations do not hold. Computing  S, T, U directly from Eqs.~(\ref{self})
would give a divergent result. However, physical results have to be
finite, so divergent quantities have to be canceled by contributions from
some higher-dimensional operators. Vector boson self energies due to scalar
loops are potentially enhanced by  $\log(\Lambda^2/m_\phi^2)$,
where $\Lambda$ provides a natural cut-off for the effective theory.
Retaining only logarithmically-enhanced parts we obtain the following
contributions to S, T and U:
\begin{eqnarray}
S & = & \frac{s_W^2 c_W^2}{2 \pi^2 \alpha}
\frac{-4 |a|^2 m_\phi^2 + |b|^2 m_Z^2}{\Lambda^2}
\log(\frac{\Lambda^2}{m_\phi^2}), \nonumber \\
T & = & 0, \\
U & = & - \frac{s_W^2}{2 \pi^2 \alpha}
\frac{4 |a|^2 m_\phi^2 + |b|^2 (m_\phi^2+s_W^2 m_Z^2)}{\Lambda^2}
\log(\frac{\Lambda^2}{m_\phi^2}). \nonumber
\end{eqnarray}
There are also threshold contributions originating from confining
effects at the scale $\Lambda$, but we neglect them.

Current limits on the S and  U parameters do not exclude large
magnitudes for the operators \ogg\ and \ogz. For instance, for a 400 GeV
scalar particle, $\Lambda = 1 {\rm TeV}$ and $a=e^2$, the contributions
from \ogg\ are only $S=-0.013$ and $U=-0.017$, while experimental bounds are
$S=-0.48 \pm 0.40$ and $U=-0.12 \pm  0.69$ \cite{STU}. The photon always
couples with strength $e$, so the values $a=e^2$ and $b=e$ are upper
bounds for the magnitudes of the operators \ogg\ and \ogz, correspondingly.

\section{Conclusions}

In this paper we have examined the decays \hgg\ and \hgz\ of the scalar
particle present in the Abbott-Farhi model. Since confining theories are not
completely understood, one can not conclusively argue for or against the
SCSM based upon the analysis of deviations from the SM. Despite the precision
of current measurements, large partial widths of the decays \hgg\ and \hgz\
are not excluded in the SCSM. The SCSM scalar particle can be discovered
at LHC by looking for \hgg\ decays as long as the scalar is lighter than
450 GeV. The observation of the decays \hgg\ and \hgz\ with partial widths
larger than predicted by the SM would indicate new, interesting physics.

\section*{Acknowledgments}
I thank Lisa Randall for suggesting this investigation and helpful remarks,
Edward Farhi and Robert L.~Jaffe for discussions. I also thank Eric Sather
for discussions and comments on the manuscript.

\newpage

\section*{Figure Captions}
\begin{itemize}
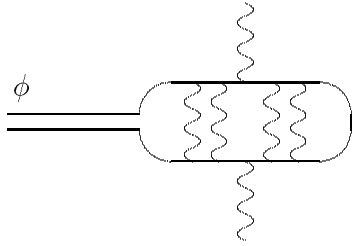
 \item[Figure 1.]
Feynman diagram contributing to the decay $\phi \rightarrow
\gamma \gamma$. The external lines represent photon fields, while the
internal lines represent strongly interacting $SU(2)_L$ bosons.
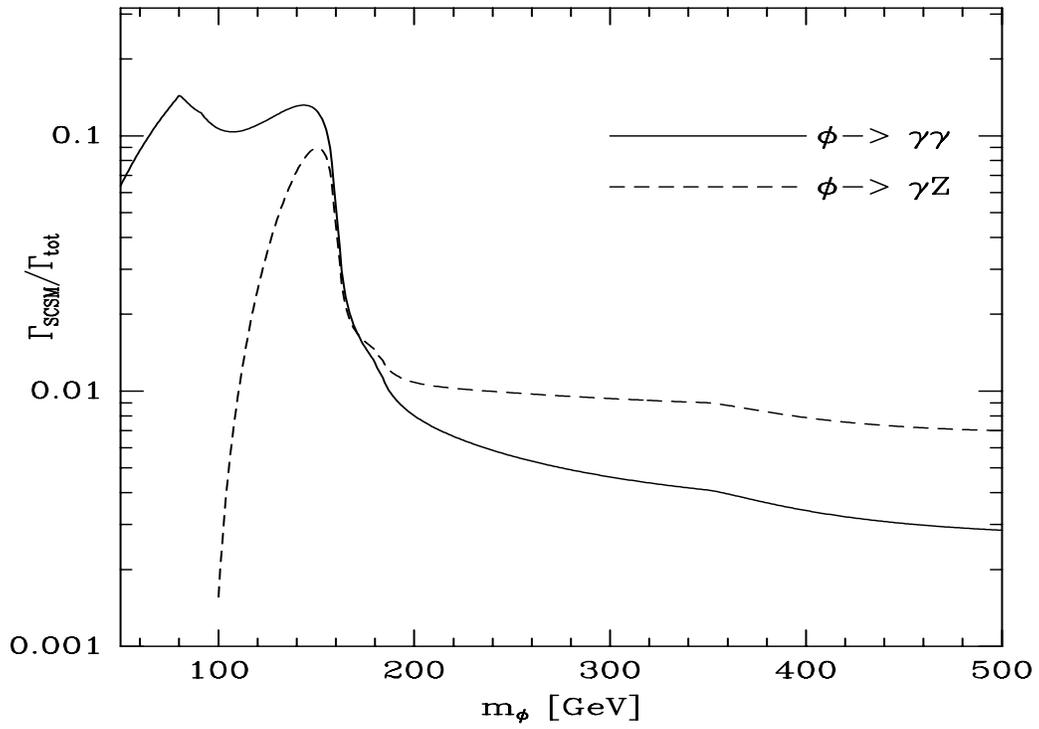
 \item[Figure 2.]
Branching ratios of the \protect\hgg\ and \protect\hgz\ channels
for $\Lambda = 1\ {\rm TeV}$.
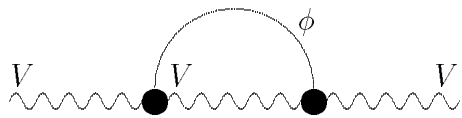
 \item[Figure 3.]
Self-energy diagram. The blobs represent $O_{\gamma \gamma}$ or
$O_{\gamma Z}$, and $V$ is either $\gamma$ or $Z$.
\end{itemize}

\section*{Table Caption}
\begin{itemize}
 \item[Table 1.] The number of events for the decays \hgg\ and \hgz\
at the LHC computed for an integrated luminosity of $10^5 \ {\rm pb}^{-1}$.
\end{itemize}

\section*{Table 1}
\begin{center}
\begin{tabular}{|l|c|c|c|c|c|} \hline
$ m_\phi$ [GeV] & 100 & 200 & 300 & 400 & 500
\\ \hline
$ \phi$ / year  & $5 \times 10^6$ & $4 \times 10^6$   &   $10^6$       &
$4 \times 10^5$ & $2 \times 10^5$ \\ \hline
\hgg            & $5 \times 10^5$ & $3.6 \times 10^4$ & $5 \times 10^3$&
$1.2 \times 10^3$ & $600$ \\ \hline
\hgz            & $5 \times 10^3$ & $4 \times 10^4$   &   $10^4$       &
$3.2 \times 10^3$ & $1.4 \times 10^3$ \\ \hline
\end{tabular}
\end{center}

\end{document}